\documentclass[printer]{aa} 
\usepackage{graphicx}
\usepackage{natbib}
\def\hmpc{h$^{-1}$ Mpc}

\def\xir{$\xi(r)$\ }

\def\xip{$\xi(r_p,\pi)$\ }
\def\wp{$w_p(r_p)$\ }

\begin{document}
   \title{The VIMOS VLT Deep Survey
         \thanks{based on data
         obtained with the European Southern Observatory Very Large
         Telescope, Paranal, Chile, program 070.A-9007(A), and on data
         obtained at the Canada-France-Hawaii Telescope, operated by
         the CNRS of France, CNRC in Canada and the University of Hawaii}
}

   \subtitle{The evolution of galaxy clustering to $z\simeq2$ from first 
epoch observations}

   \author{
O. Le F\`evre \inst{1}, 
L. Guzzo \inst{2}, 
B. Meneux \inst{1},
A. Pollo \inst{2}, 
A. Cappi \inst{3}, 
S. Colombi \inst{9},
A. Iovino \inst{2}, 
C. Marinoni \inst{1,2},
H.J. McCracken \inst{9,12}, 
R. Scaramella \inst{7}, 
D. Bottini \inst{4}, 
B. Garilli \inst{4}, 
V. Le Brun \inst{1}, 
D. Maccagni \inst{4}, 
J.P. Picat \inst{5},  
M. Scodeggio \inst{4}, 
L. Tresse  \inst{1},
G. Vettolani \inst{6},
A. Zanichelli \inst{6}, 
C. Adami \inst{1}, 
M. Arnaboldi \inst{11},
S. Arnouts \inst{1},
S. Bardelli \inst{3},
J. Blaizot \inst{1},
M. Bolzonella \inst{8},
S. Charlot \inst{9,10}, 
P. Ciliegi\inst{6}, 
T. Contini \inst{5},
S. Foucaud \inst{4},  
P. Franzetti \inst{4},
I. Gavignaud \inst{5,12}, 
O. Ilbert \inst{1,3}, 
B. Marano \inst{8}, 
G. Mathez \inst{5}, 
A. Mazure \inst{1},
R. Merighi \inst{3}, 
S. Paltani \inst{1},
R. Pell\`o \inst{5}, 
L. Pozzetti \inst{3},
M. Radovich \inst{11}, 
G. Zamorani  \inst{3}, 
E. Zucca  \inst{3},
M. Bondi \inst{6}, 
A. Bongiorno \inst{8}
G. Busarello \inst{11}, 
Y. Mellier \inst{9,12}, 
P. Merluzzi \inst{11}, 
V. Ripepi \inst{11},
D. Rizzo \inst{5,2} 
          }

   \offprints{O. Le F\`evre}

   \institute{
Laboratoire d'Astrophysique de Marseille, UMR 6110 CNRS-Universit\'e
de Provence, Traverse 
    du Siphon - B.P.8, F-13012 Marseille, France\\
              email: olivier.lefevre@oamp.fr
         \and
INAF - Osservatorio Astronomico di Brera, via Brera 28, Milan, Italy
\and
INAF - Osservatorio Astronomico di Bologna, via Ranzani 1, 40127 Bologna, Italy
\and 
INAF - IASF, via Bassini, 15, I-20133, Milano, Italy
\and
Laboratoire d'Astrophysique de l'Observatoire Midi-Pyr\'en\'ees, UMR5572, 
14 Av. Ed. Belin, F-31400 Toulouse, France
\and
INAF - Istituto di Radio-Astronomia, via Gobetti 101, I-40129, Bologna, Italy
\and
INAF - Osservatorio Astronomico di Roma, Italy
\and
Dipartimento di Astronomia, Universit\`a di Bologna, via Ranzani 1, I-40127 Bologna, Italy
via Ranzani 1, I-40127 Bologna, Italy
\and
Institut d'Astrophysique de Paris, UMR 7095, 98 bis Bvd Arago, F-75014 Paris, France
\and
Max Planck Institut fur Astrophysik, D-85741 Garching, Germany
\and
INAF - Osservatorio Astronomico di Capodimonte, via Moiariello 16, I-80131 Napoli, Italy
\and
Observatoire de Paris, LERMA, UMR 8112, 61 Av. de l'Observatoire, F-75014 Paris, France
\and
European Southern Observatory, Karl Schwarzschild Str. 2, D-85748 Garching
bei Munchen, Germany
             }

   \date{Received September 7, 2004; accepted April 11, 2005}

   \abstract{This paper presents the evolution of the
clustering of the main population of 
galaxies from $z\simeq2$ to $z=0.2$, from the first epoch 
VIMOS VLT Deep Survey (VVDS), a magnitude limited sample with 
$17.5 \leq I_{AB} \leq 24$. 
The sample allows a direct estimate of evolution {\it
from within the same survey} over the time base sampled.
We have computed the correlation functions
$\xi(r_p,\pi)$ 
and $w_p(r_p)$,
and the correlation length $r_0(z)$,
for the VVDS-02h and VVDS-CDFS fields,
for a total of 7155 galaxies in a
$0.61$ deg$^2$ area. We find that the correlation length
in this sample slightly increases from $z=0.5$ to $z=1.1$,
with $r_0(z)=2.2-2.9$ h$^{-1}$ Mpc (comoving), for galaxies
comparable in luminosity to the local 2dFGRS 
and SDSS samples, 
indicating that the amplitude of the correlation function
was $\simeq2.5$ times lower at $z\simeq1$ than observed
locally. The correlation length in our lowest redshift bin $z=[0.2,0.5]$
is $r_0=2.2$ h$^{-1}$ Mpc, lower than for any other population at 
the same redshift, indicating the low clustering of
very low luminosity galaxies, 1.5 magnitudes fainter than in the
2dFGRS or SDSS. The correlation length
increases to $r_0\sim3.6$ h$^{-1}$ Mpc at higher redshifts $z=[1.3,2.1]$,
as we are observing increasingly brighter galaxies, comparable
to galaxies with $M_{B_{AB}}=-20.5$ locally.
We compare our measurement to the DEEP2 measurements in 
the range $z=[0.7,1.35]$ \citep{coil} and find comparable results
when applying the same magnitude and color selection 
criteria as in their survey.
The slowly varying clustering of VVDS galaxies as redshift increases
is markedly different from the predicted evolution 
of the clustering of dark matter,
indicating that bright galaxies traced higher density
peaks when the large scale structures were emerging from the
dark matter distribution $9-10$ billion years ago,
being supporting evidence for a strong evolution of the 
galaxy vs. dark matter bias. 
   }

\authorrunning{Le F\`evre, O., Guzzo, L., et al.}
\titlerunning{VVDS: clustering evolution from $z\simeq2$}

   \maketitle
%

\section{Introduction}

The evolution of the clustering of galaxies is a key diagnostic 
element to test the evolution of the universe, allowing direct
comparison between observations and theory. 
In the current paradigm of galaxy formation and evolution,
dark matter halos that contain galaxies are expected to merge and grow 
under the action of gravity. This translates into a
continuous evolution of the correlation function $\xi(r,z)$ of
dark matter halos, now quite well understood from extensive 
high resolution numerical simulations 
\citep[see][]{weinberg,benson,somerville,kauffmann}. As a direct 
measurement of the space distribution
of dark matter halos 
%
is not yet feasible, we are compelled to use  
galaxies as indirect tracers of the dark matter. Unfortunately, 
as galaxies are complex physical systems, their relationship
to
the underlying mass,
the ``bias'', is difficult to estimate.  
As galaxies and dark matter evolve, the bias may evolve and relating the
measurements of galaxy clustering to the evolution of the total mass 
is not easy, with the bias
shown to depend upon galaxy type, luminosity
and local environment \citep{norberg2}.

The most straigthforward indicator of galaxy clustering is 
the correlation function $\xi(r)$, representing the excess
probability over random
of finding a galaxy in a given volume,
at a fixed distance from another galaxy.  
The shape and amplitude of the galaxy correlation function at the
current epoch is now established to high accuracy.  $\xi(r)$ is well
described by a power law $\xi(r)=(r/r_0)^{-\gamma}$ over scales
$0.1-10$h$^{-1}$ Mpc \citep{dp,hawkins,zehavi04}, 
with a more refined
modelling requiring some extra power over this shape for separations
larger than 2 - 3 \hmpc\ \citep{guzzo91,zehavi04}, 
a feature possibly encoding information on the
relation between galaxies and their host dark-matter halos.
The local clustering measurements have shown that the
correlation length $r_0$ increases from late types to early type
galaxies, from low luminosity to high luminosity and from low to high
galaxy density environments \citep[see][]{giovanelli,benoist,
guzzo97,norberg}, with luminosity being the
dominant effect \citep{norberg2}.  
The most recent estimates of the 
correlation length from the 2dFGRS and SDSS vary from $r_0=3$ for late type
star-forming galaxies in low density environments to $r_0=5-6$ for
galaxies with $M_*=-19.5$, with the clustering amplitude 
reaching $r_o=7.5$ h$^{-1}$ Mpc for galaxies four times $L_*$. 

At higher redshifts
the situation is less clear.
Analysis of the projected angular correlation function favors
a stable clustering \citep{postman,roche,mccracken01,cabanac} 
but it requires a priori knowledge of the redshift distribution
of the galaxy population sampled.
A variety of results have been obtained from smaller
spectroscopic samples, with comoving
correlation lengths $r_0$ in the range $2-5$ \hmpc\ at $z\sim0.5$
\citep{lefevre96,small,sheperd,carlberg}
and $r_0=3-5$ \hmpc\ at $z\sim3$ \citep{adel,foucaud}.  
Recently, first results from the DEEP2 survey have been presented,
indicating a correlation length   
$r_0=3.53\pm0.81$ h$^{-1}$ Mpc in z=[0.7,0.9], and 
$r_0=3.12\pm0.72$ h$^{-1}$ Mpc in z=[0.9,1.35]
\citep{coil}.
The main difficulty in interpreting these results in terms of
evolution of the clustering is to relate the
population of galaxies observed at a given redshift to a lower 
redshift population of ``descendants'', identified from
a well-defined selection function enabling comparisons.  
Some of these surveys are targeted to specific classes of
galaxies, pre-selected via photometric methods, whose relation to the
global population is not obvious.  The most notable example is
represented by galaxies selected via the Lyman-break technique around
$z\sim 3$ and $z\sim 4$ \citep{Steidel98}, which
display a clustering strength similar to present-day normal galaxies
and therefore represent a very {\it biased} population, possibly the
precursors of giant cluster ellipticals \citep{governato}.

On the other hand, even when selecting purely magnitude-limited samples, one 
cannot avoid being affected by the complex dependence of clustering on
morphology and luminosity
evidenced by the
%
wide range of correlation lengths measured in the local Universe, 
and its evolution as a function of redshift.
One needs to observe samples
at increasingly high redshifts with the same luminosities, colors
(type) and local environments in order to derive the evolution of the
clustering of galaxies and hence attempt to derive how the
correlation properties of the mass evolve. At high redshifts, the
natural observational bias is to sample increasingly brighter and more
actively star forming galaxies, which may have a direct impact on our
current vision of the evolution.

Finally, high-redshift samples of spectroscopically measured 
galaxies have been
inevitably limited so far to relatively bright objects
in small areas on the sky, which
contributes to increase the scatter between independent measurements,
further complicating their interpretation.


In this paper we present the first attempt to measure
the evolution of the clustering 
in a consistant way accross the redshift range $0.2 < z \leq 2.1$,
using 7155 galaxies from the VIMOS VLT Deep Survey (VVDS)
over more than $0.61$ deg$^2$.
The VVDS is designed to sample the
high redshift population of galaxies in the most unbiased
way possible, using a simple
magnitude selection in the range $17.5 \leq I_{AB} \leq 24$,
using several independant fields up to 4 deg$^2$ each \citep{lefevre04c}.
This analysis uses the high-quality First Epoch VVDS sample,
which includes 6117 galaxy redshifts in the VVDS-02h
\citep{lefevre04c} and another 1368 galaxies in the VVDS-CDFS 
\citep{lefevre04b} field. We have measured the redshift-space
correlation function $\xi(r_p,\pi)$, and computed the projected
function $w_p(r_p)$, to recover the value of the
galaxy correlation length $r_0(z)$  up to $z\simeq2$,
therefore tracing the evolution of the clustering over
more than 10 Gy, or 70\% of the current age of the universe.

In section 2 we recall the properties of the VVDS First Epoch sample.
In section 3, we describe how the correlation function has been
computed, referring in large part to the accompanying paper by \citet{pollo1}
which describes all the methods set up to validate the measurements
and compute the errors. In section 4, we present the results
in terms of the evolution of $r_0(z)$, and
we compare our results to previous measurements where possible. 
In section 5, we discuss the evolution of the clustering of
the global population of galaxies from $z\simeq2$, before concluding
in section 6. 

This paper is the first in a series of papers to study the
clustering of galaxies at high redshift from the VVDS
first epoch data. Guzzo et al. (2005, in prep.)
will present the clustering evolution from volume limited
samples and infer the dependence of clustering
upon luminosity; Meneux et al. (2005, in prep.) will present the differences 
in clustering observed as a function of galaxy type and its
evolution and Pollo et al. (2005, in prep.)
will investigate the dependence and evolution of clustering as a function 
of the local environment. \citet{marinoni} and Le F\`evre et al.
(2005, in prep.) will look
at the evolution of the galaxy -- dark matter 
bias, and subsequent papers will study the 
clustering from the redshift population $2 \leq z \leq 5$.


We have used a Concordance Cosmology with $\Omega_m=0.3$, and
$\Omega_{\Lambda}=0.7$ throughout this paper.  The Hubble constant is
normally parameterized via h$=$H$_\circ/100$, to ease comparison to
previous works, while a value H$_\circ=70$ km s$^{-1}$ Mpc$^{-1}$ has
been used when computing absolute magnitudes.  All correlation
length values are quoted in comoving coordinates.

\section{VVDS first epoch data}

\subsection{The sample}

\begin{figure}
   \centering
      \includegraphics[width=8cm]{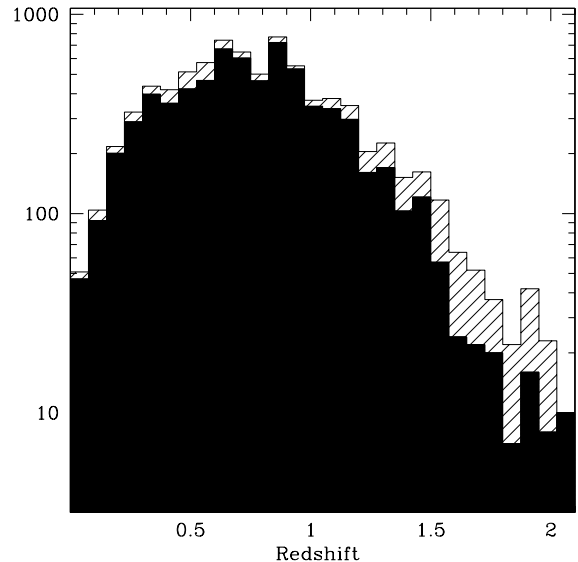}
      \caption{The redshift distribution in the VVDS-02h field. The
filled histogram contains 6137 galaxies with quality flag $\geq2$,
and another 1093 galaxies with quality flag $=1$ (open
dashed histogram), in the redshift
range z=[0.2,2.1]
              }
         \label{histz}
   \end{figure}

The VVDS-Deep sample is stricly selected in magnitude in the 
range $17.5 \leq I_{AB} \leq 24$, from a complete deep
photometric survey \citep{lefevre04a,mccracken03} 
without any color or
shape selection \citep{lefevre04b}.
We have analysed two fields, the VVDS-02h and the VVDS-CDFS. 
Over the $\sim9600$ redshifts measured in these two fields, 
we have used in the following analysis 
only objects with a redshift confidence level 
higher than or equal to $\simeq80$\% (quality flags 2 to 9
as defined in \cite{lefevre04b}),
excluding QSOs.  We will only mention briefly below 
the effect on measured
correlations of relaxing the quality threshold, including the
poorest redshift measurements (flag 1). 
The complete sample analysed concerns a total of 7155 galaxies
in $2203$ arcmin$^2$, with 6137 galaxies in 
the $1750$ arcmin$^2$
VVDS-02h field with $0.2 \leq z \leq 2.1$, and 1038 galaxies 
in the $453$ arcmin$^2$ VVDS-CDFS
(Chandra Deep Field South, \citep{gia}) area with $0.4 \leq z \leq 1.5$. 
The accuracy of the redshift measurements 
is $\simeq275$ km/s \citep{lefevre04a}. The distribution of 
redshifts for the VVDS-02h field is presented in Figure~\ref{histz}. 
The redshift distribution peaks at $z\simeq0.8$, and there 
are 300 galaxies with $1.3 \leq z \leq 1.5$, and 132 galaxies
with $1.5 \leq z \leq 2.1$.

\subsection{The galaxy population mix}

\begin{figure*}
   \centering
      \includegraphics[width=15cm]{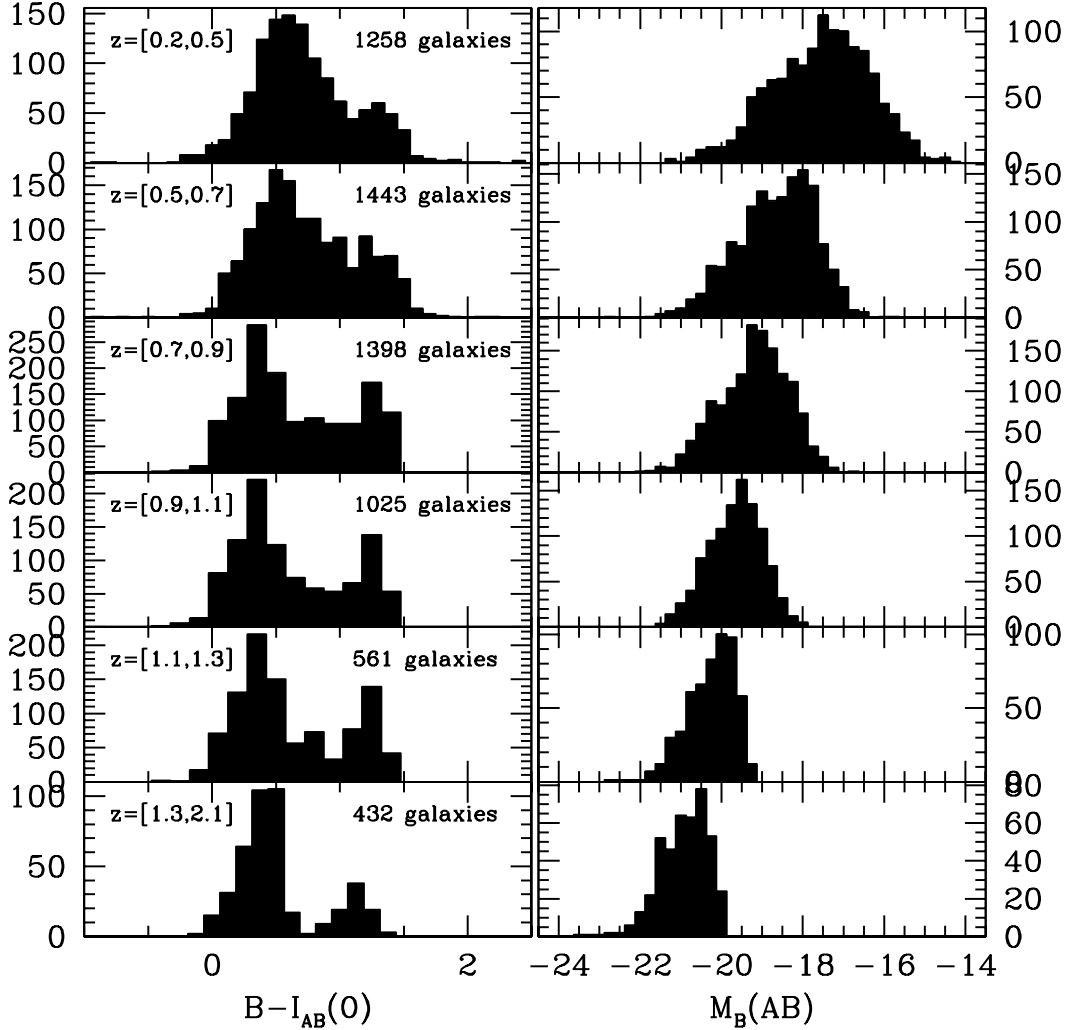}
      \caption{The rest frame $B-I_{AB}$ color (left) and 
absolute magnitude $M_{B_{AB}}$ (right) from the VVDS-02h data up 
to $z\sim2$.  
              }
         \label{MBz}
   \end{figure*}

We have split the sample into 6 redshift bins, as described in
Table~\ref{meas}. The rest frame $B-I_{AB}(0)$ color and absolute magnitude 
$M_{B_{AB}}$ distribution 
within each bin
is shown in Figure~\ref{MBz}, from
$z=0.2$ to $z=2.1$, and the mean values
are reported in Table~\ref{meas}. $B-I_{AB}(0)$ and $M_{B_{AB}}$ have
been computed using template fitting of the
photometric spectral energy distribution in
B,V,R, and I bands, to derive the $k(z)$ corrections
\citep[see][for details]{ilbert}.

Up to redshift
$z=1.3$, the $B-I_{AB}(0)$ color distribution stays
quite similar with increasing redshift, 
from blue star forming $B-I_{AB}(0)=0$
to red $B-I_{AB}(0)\sim3$,  while for $1.3<z\leq2.1$, the
reddest $B-I_{AB}(0) \geq 2.5$ galaxies, if 
present, are not observed. 
The magnitude selection of the VVDS-Deep,
therefore, allows us to sample the global population of galaxies, 
for all galaxy types from late to early types, up to $z\sim1.3$,
while for $z>1.3$, as we are selecting galaxies from their UV
rest frame continuum at these redshifts, the VVDS is increasingly 
biased towards late-type, higher star-forming galaxies. 
We therefore expect that in the farthest redshift bin,
the clustering measurement in the VVDS is the result
of the effects of looking at intrinsicaly more luminous 
and more actively star forming galaxies.

The range of absolute $M_{B_{AB}}$ magnitudes sampled
is quite large, and changes strongly with redshift
as shown in Figure \ref{MBz}. While at redshift $z\sim0.5$,
the absolute magnitude range sampled is $-22 \leq M_{B_{AB}} \leq -16$,
only bright galaxies with $M_{B_{AB}} \leq -20$ are sampled
at $z>1.3$. The absolute luminosity of galaxies
in the VVDS is shown to increase as redshift increases,
as computed from the Luminosity Function \citep{ilbert}. 
If one assumes a pure luminosity evolution, 
galaxies observed at high redshift
are expected to have faded to fainter absolute magnitudes
at the present time, by as much as $M_B=1.5$ to $2$  magnitudes. This should be
taken into account when comparing the clustering of the
high redshift population to local populations.

The consequences on the correlation function measurements
of the change in the population sampled as a function of
redshift are discussed in Section 5.

\section{Computing the real-space correlation and correlation length $r_0(z)$}

The methods applied on the VVDS first epoch data to
derive the real space correlation parameters are
described extensively in the accompanying paper
by \citet{pollo1}. We summarize below the main elements of this
method. 

\subsection{Estimating correlation functions from the VVDS}

To measure the galaxy real-space correlation length $r_\circ $ and
slope $\gamma$ from our survey, we have used the projection of the 
bi-dimensional correlation function \xip.  This function was estimated
using the well-known \citet{LandySzalay} estimator
\small
\begin{equation}
\xi(r_p,\pi) = \frac{N_R(N_R-1)}{N_G(N_G-1)} \frac{GG(r_p,\pi)}{RR(r_p,\pi)} 
	- 2 \frac{N_R-1}{N_G} \frac{GR(r_p,\pi)}{RR(r_p,\pi)} + 1
\end{equation}
\normalsize
where $N_G$ is the mean galaxy density (or, equivalently, the total number
of objects) in the survey; $N_R$ is the mean density of a catalogue of random
points distributed within the same volume of the considered redshift bin;   
$GG(r)$ is the number 
of independent galaxy-galaxy pairs with separation between $r$ and $r+dr$; 
$RR(r)$ is the number of independent random-random pairs and $GR(r)$
the number of cross galaxy-random pairs within the same 
interval of separations .   A total of $\sim40,000$ random points
have been used in each redshift bin, guaranteeing a sufficient density to
avoid shot-noise effects on small scales. The random sample follows
exactly the same geometry, redshift distribution 
and observational pattern as in the galaxy
data, while a specific weighting scheme is used to overcome the biases
introduced by the slit-positioning software and other selection
effects.  These techniques have been extensively tested on a large
number of mock VVDS surveys and have been shown to be able to recover
the correct $w_p(r_p)$ correlation function 
to better than $10\%$, reducing any systematic effect
to less than 5\%. We have not applied this
small correction to the data presented in this paper.
This is discussed in detail in
the accompanying paper by \citet{pollo1}.

Since we are not computing the correlation function from the full
magnitude-limited survey altogether, there is no point 
here in using the so-called
$J_3$ minimum-variance weighting.  This usually has been adopted in
the analysis of large flux-limited local surveys
\citep{Fisher94,esp00}, in which the sampling of the
clustering process varies dramatically between the nearby and distant
parts of the sample. Its main scope is to avoid excessive weighting of
the most distant parts of the sample, where only sparse bright
galaxies trace structures.  Within each of our redshift bins,
the density of objects varies only slightly and equal weighting of the
pairs is the most appropriate choice \citep{Fisher94}.

\subsection{$\xi(r_p,\pi)$, $w_p(r_p)$ and the correlation length $r_0$}

We have computed the two point correlation function $\xi(r_p,\pi)$ 
in increasing redshift bins, selecting the bin boundaries to
maximize the number of objects, hence the signal to noise of the
correlation function, in each of the bins.  

The formalism developed by \citet{dp} has been used to derive
the real space correlation function in the presence of 
redshift-space distortions along the line of sight. 
We integrate $\xi(r_p,\pi)$ along the line of sight,
computing 

\begin{equation}
w_p(r_p) \equiv 2 \int_0^\infty dy \xi(r_p,\pi) = 2 \int_0^\infty dy 
\xi\left[(r_p^2+y^2)^{1/2}\right],
\label{wpdef}
\end{equation}
where in practice the upper integration limit  has to be chosen finite, 
to avoid adding noise to the result.  After a few experiments, we 
used a value of $20$ h$^{-1}$ Mpc.   

If we assume a power-law form for $\xi(r)$, i.e. 

\begin{equation}
\xi(r) = \left(\frac{r}{r_0}\right)^{-\gamma},
\label{powlaw}
\end{equation}
$w_p(r_p)$ can be written as

\begin{equation}
w_p(r_p) = r_p \left(\frac{r_0}{r_p}\right)^\gamma 
\frac{\Gamma\left(\frac{1}{2}\right)\Gamma\left(\frac{\gamma-1}{2}\right)}
{\Gamma\left(\frac{\gamma}{2}\right)},
\label{wpmodel}
\end{equation}
where $\Gamma$ is the Euler Gamma function. 

Fitting the $w_p(r_p)$ measurements in each redshift bin
then provides a measurement of $r_0(z)$ and $\gamma(z)$.

\subsection{Computing errors}
\label{errors}

The uncertainty associated with the computation of $r_0(z)$ and 
$\gamma(z)$ is largely dominated by cosmic variance. Although
our spectroscopic sample is the largest available to date at the
redshifts probed, both in the number of galaxies
and area surveyed, only two fields have been sampled,
with the VVDS-02h largely dominating over the VVDS-CDFS in
terms of the number of galaxies observed and area covered,
and it is therefore 
inappropriate
to estimate errors
on $w_p(r_p)$, $r_0(z)$, and $\gamma(z)$ directly from
field-to-field variations within our data set.  
Instead, we have been using 
 ensemble errors derived from the 
scatter in the correlation function computed from 50 
mock VVDS surveys constructed using the GalICS simulations 
\citep{blaizot}, as thoroughly described in the parallel
paper by  \citep{pollo1}.
These realistic mock samples specifically include the same
selection function and observational biases present in the
VVDS First Epoch data set, and allow us to compute 
ensemble statistical averages and scatters of the quantities 
we measure on the real data. In particular, the standard 
deviation of the measured $w_p(r_p)$ from the mock VVDS samples
provides us with a realistic set of error bars 
that include also the effect of 
field-to-field variations due to fluctuations on scales larger than the 
observed field, i.e. cosmic variance (obviously assuming that the 
$\lambda$-CDM power spectrum and normalization are a good match to the 
observed one, which is indeed the case).

The comparison of the VVDS-02h and VVDS-CDFS 
correlation functions gives us an
external check to this procedure, with a
relatively noisy indication of the amplitude of cosmic variance,
as described below.

\section{Results}

\subsection{The VVDS correlation function
from a magnitude limited sample $17.5 \leq I_{AB} \leq 24$ out
to $z\simeq2$}

We have computed the correlation function $\xi(r_p,\pi)$
and its projection $w_p(r_p)$ on both the VVDS-02h and 
VVDS-CDFS fields. 

\begin{figure*}
   \centering
      \includegraphics[width=18cm]{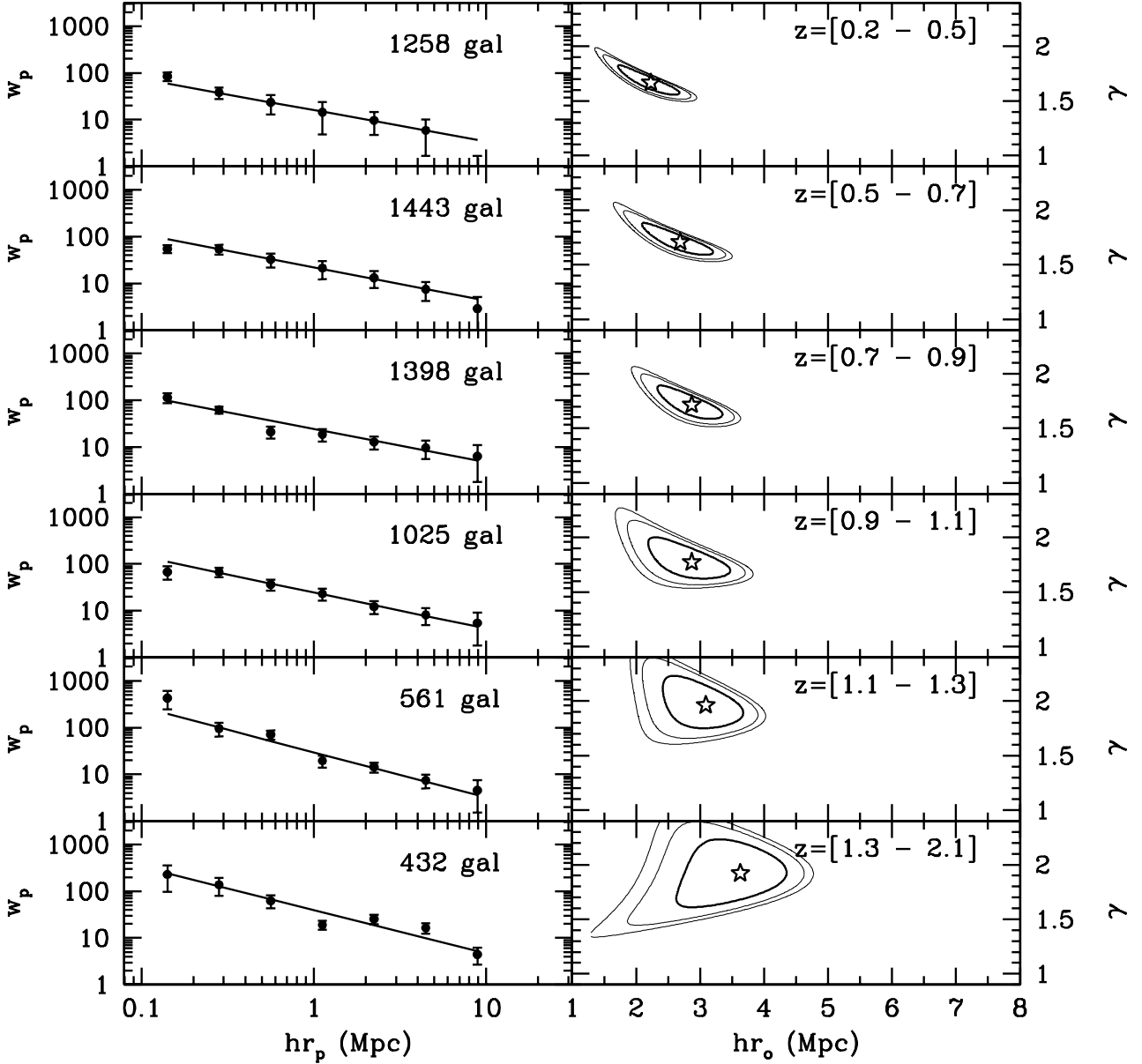}
      \caption{The correlation function from the VVDS-02h data up 
to $z\simeq2$: (left) $w_p(r_p)$ measured in 6 redshift
bins, the number of galaxies in each bin is indicated
on the top right of each panel, (right) the correlation
length $r_0(z)$ and correlation function slope $\gamma$
measured from fitting $w_p(r_p)$
assuming a power law form  $\xi(r) = (r/r_0)^{-\gamma}$. 
$68$\%, $90$\% and $95$\% likelihood contours
are drawn.
}
         \label{cor02}
   \end{figure*}

\begin{figure*}
   \centering
      \includegraphics[width=9cm]{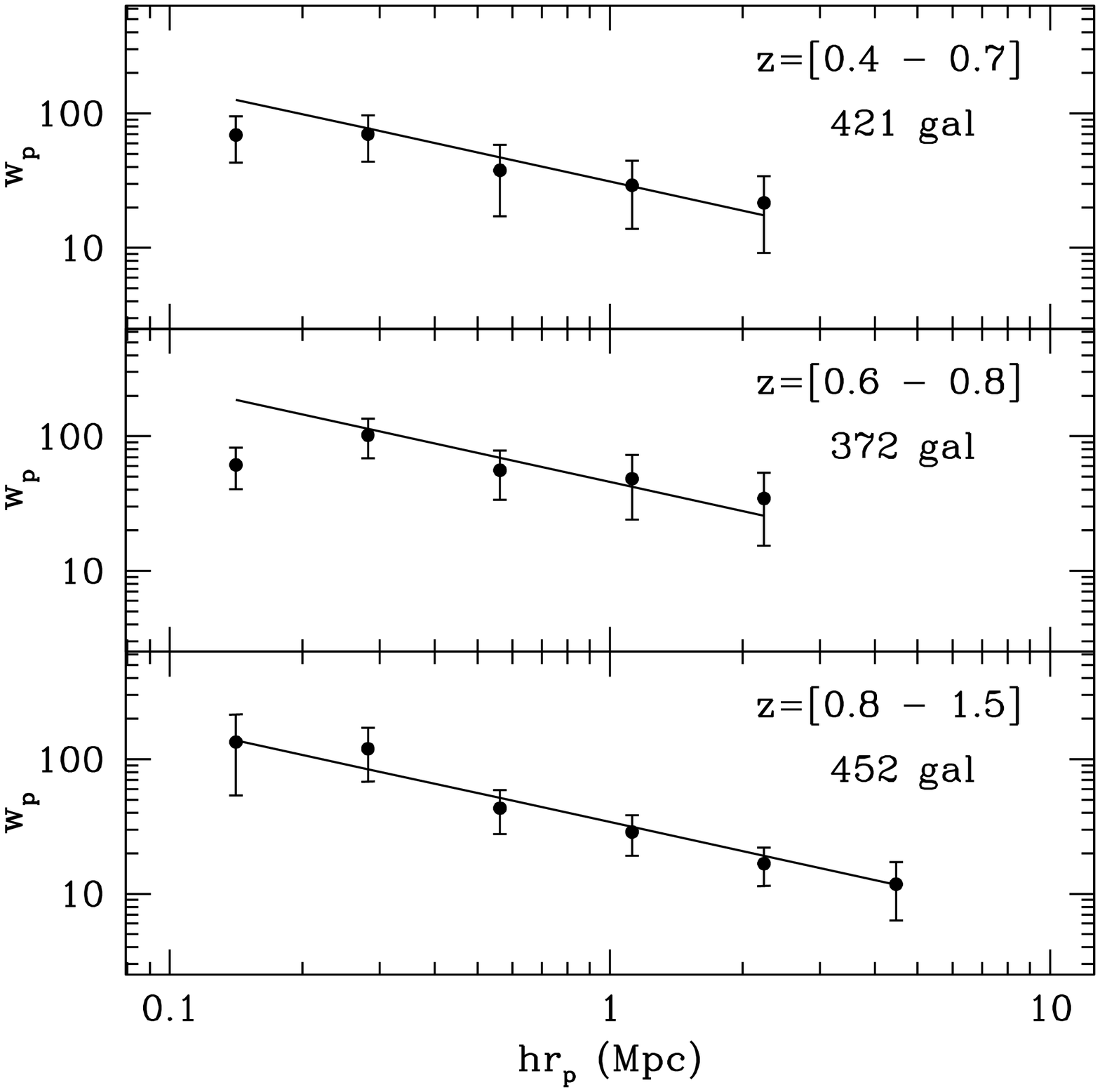}      
\caption{The correlation function $w_p(r_p)$ for the VVDS-CDFS data in
three redshift bins from $z=0.4$ to 
to $z=1.5$. The slope has been fixed to $\gamma=1.72$, i.e. the mean
measured in the VVDS-02h field in the range z=[0.2,1.1] 
              }
         \label{corrcdfs}
   \end{figure*}

The correlation function $w_p(r_p)$ is presented in the
left panel of Figure
\ref{cor02} for the VVDS-02h field, and in Figure \ref{corrcdfs}
for the VVDS-CDFS field. The number of galaxies observed
in each field is indicated for each redshift bin
in Figure \ref{cor02}, so far the largest sample of galaxies 
used to compute the correlation function at these redshifts.
The large sample allows to compute $w_p(r_p)$ in
6 redshift bins up to $z=2.1$ for the VVDS-02h field,
and in 3 bins up to $z=1.5$ for the VVDS-CDFS field.
Error bars are ensemble errors computed 
as described in \S~\ref{errors}.   
A positive correlation signal is measured 
out to at least 30 \hmpc\ in all bins, and \wp is well described
by a power law in the range $0.1 \leq r_p \leq 10$ h$^{-1}$ Mpc (note
however that any redshift space feature 
in \xir is smoothed out in \wp, which is its integral).  
The measured correlation function
amplitude is relatively low at low redshifts $z \leq 0.5$, and
stays essentially constant as a function of redshift, with just a
mild increase  in the farthest bins.
A possible interpretation of these results is discussed in the next sections.

\subsection{Evolution of the galaxy real-space correlation function}

The measured values of the correlation length
$r_0(z)$ and the correlation function slope
$\gamma(z)$ computed from the fitting of
$w_p(r_p)$ are reported in Table \ref{meas}. For the 
VVDS-02h field, we have used all $w_p$
points for $0.1 \leq r_p \leq 10$ h$^{-1}$ Mpc.
For the VVDS-CDFS field, we have used points in the range 
$0.1 \leq r_p \leq 3$ h$^{-1}$ Mpc only, because of the smaller
field size. We also report in Table \ref{meas} the values of
$r_0(z)$ obtained after fitting $w_p$ with the slope $\gamma$
fixed to the average slope measured in the range z=[0.2,1.3].
In the VVDS-CDFS, a strong wall-like 
large scale structure has been identified 
at $z\sim0.735$, with more than 130 galaxies
in a velocity range $\pm2000$ km/s \citep{lefevre04b}, and is
expected to strongly affect the correlation function
computation. The correlation
function in this field has been computed 
in the redshift interval z=[0.6,0.8] which
includes this strong structure, as well as in z=[0.4,0.7]
and z=[0.8,1.5] to specifically exclude it, 
as reported in Table~\ref{meas}.
While we find a slightly higher correlation length in the
VVDS-CDFS, the measurements are compatible, within
the errors, with the values reported for the VVDS-02h. 

We have investigated how our results change when
using the full set of available redshifts, i.e. including also the
1300 less accurate measurements (flag 1), 
which are shown to be $\sim$55\% correct \citep{lefevre04c}.  
The result is that the measured
correlation lengths are lowered by $\sim3-5\%$, which we interpret as 
a consequence of the significant fraction of poorly measured redshifts 
which dilute the actual correlation function.

The results strikingly show a slowly increasing correlation length
over the complete redshift range z=[0.2,2.1]. The lowest value
measured is in the lowest redshift bin probed, then $r_0$
is constant $r_0 \sim 2.8$ h$^{-1}$ Mpc over the range
z=[0.5,1.1] and increases slightly to $r_0 \sim 3.6$ h$^{-1}$ Mpc
over z=[1.1,2.1]. When we fit the slope $\gamma$ of the correlation function
at the same time as $r_0$, it varies between $1.67$ and $1.96$, slightly
increasing with redshift, as reported 
in Table \ref{meas}. The average slope in the range
z=[0.2,1.3] is $\gamma=1.76$.

   \begin{table*}
     \centering
      \caption{Measurements of the correlation length $r_0(z)$
and the correlation function slope $\gamma(z)$. The associated
$1\sigma$ errors are reported. $r_0$ values are computed
both with letting $\gamma$ free, and setting $\gamma$ to
the mean $1.78$ in the range z=[0.2,1.3] in the VVDS-F02 and VVDS-CDFS.
To compare to the DEEP2 measurements of \citet{coil},
we have set $\gamma=1.66$ for the VVDS-F02-DEEP2 in z=[0.7,0.9] and
z=[0.9,1.35], and $\gamma=1.68$ for the VVDS-F02-DEEP2 in z=[0.7,1.35].
The redshift range,
number of galaxies used, mean absolute magnitude $M_{B_{AB}}$ and
rest frame color $B-I_{AB}(z=0)$, are indicated for each subsample.}
      \[
        \begin{array}{llcccccc}
           \hline
            \noalign{\smallskip}
            $Field$ & $Redshift$ & N_{gal} & mean       & mean          & r_0(z) & \gamma & r_0(z)\\
                    &  range     &         & M_{B_{AB}} & B-I_{AB}(z=0) & h^{-1} Mpc  &        & ($fixed$\ \gamma) \\     
            \noalign{\smallskip}
            \hline
            \noalign{\smallskip}
            VVDS-02h & [0.2-0.5] &  1258   &     -17.57 &   1.34 & 2.23_{-0.51}^{+0.46} & 1.67_{-0.11}^{+0.17} &  2.27_{-0.52}^{+0.47}\\
            \noalign{\smallskip}
            \noalign{\smallskip}
            VVDS-02h & [0.5-0.7] &  1443   &     -18.68 &   1.28 & 2.69_{-0.59}^{+0.53} & 1.71_{-0.11}^{+0.18} &  2.51_{-0.55}^{+0.49}\\
            \noalign{\smallskip}
            \noalign{\smallskip}
            VVDS-02h & [0.7-0.9] &  1398   &     -19.20 &   1.36 & 2.87_{-0.52}^{+0.51} & 1.72_{-0.12}^{+0.19} &  2.87_{-0.52}^{+0.50}\\
            \noalign{\smallskip}
            \noalign{\smallskip}
            VVDS-02h & [0.9-1.1] &  1025   &     -19.66 &   1.35 & 2.87_{-0.66}^{+0.62} & 1.77_{-0.14}^{+0.24} &  2.75_{-0.63}^{+0.60}\\
            \noalign{\smallskip}
            \noalign{\smallskip}
            VVDS-02h & [1.1-1.3] &  561    &     -20.26 &   1.35 & 3.09_{-0.65}^{+0.61} & 1.96_{-0.21}^{+0.27} &  2.93_{-0.62}^{+0.58}\\
            \noalign{\smallskip}
            \noalign{\smallskip}
            VVDS-02h & [1.3-2.1] &  432    &     -20.93 &   1.25 & 3.63_{-0.76}^{+0.63} & 1.92_{-0.31}^{+0.31} &  3.69_{-1.00}^{+0.77}\\
            \noalign{\smallskip}
            \noalign{\smallskip}
            VVDS-02h $``DEEP2'' selection$ & [0.7-1.35] &  1620 & -20.00 & 1.36 &  3.05_{-0.53}^{+0.51} & 1.56_{-0.11}^{+0.14} & 3.17_{-0.55}^{+0.53} \\
            \noalign{\smallskip}
            \noalign{\smallskip}
            VVDS-02h $``DEEP2'' selection$ & [0.7-0.9] &  687 & -19.65 & 1.41 & 3.59_{-0.83}^{+0.77} & 1.65_{-0.17}^{+0.25} & 3.59_{-0.83}^{+0.77} \\
            \noalign{\smallskip}
            \noalign{\smallskip}
            VVDS-02h $``DEEP2'' selection$ & [0.9-1.35] &  933 & -20.25 & 1.32 &  3.29_{-0.59}^{+0.57} & 1.82_{-0.15}^{+0.19} & 3.13_{-0.57}^{+0.54} \\
            \noalign{\smallskip}
            \noalign{\smallskip}
              VVDS-CDFS & [0.4-0.7] &  421 & -19.44 & 1.30 & 3.19_{-1.01}^{+0.93} & 1.45_{-0.17}^{+0.25} & 3.31_{-1.26}^{+1.06} \\
            \noalign{\smallskip}
            \noalign{\smallskip}
            VVDS-CDFS & [0.6-0.8] & 372 & -19.74 & 1.47 & 4.55_{-1.46}^{+1.25} & 1.48_{-0.15}^{+0.28} & 4.15_{-1.50}^{+1.25}\\
            \noalign{\smallskip}
            \noalign{\smallskip}
            VVDS-CDFS & [0.8-1.5] & 452 & -20.43 & 1.34 & 3.53_{-1.66}^{+1.29} & 1.66_{-0.27}^{+0.40} & 3.49_{-1.78}^{+1.28}\\
            \noalign{\smallskip}
            \hline
         \end{array}
      \]
\label{meas}
   \end{table*}

\subsection{Comparison with other surveys}

In redshift surveys of the local Universe ($z<0.2$), the lowest values
for the correlation length have been measured for late-type 
star forming galaxies as e.g. H-II galaxies ($r_0=2.7$ h$^{-1}$ Mpc,
\citet{H2gals}) and  infrared-selected {\it IRAS} galaxies
($r_0=3.76$ h$^{-1}$ Mpc, \citet{Fisher94}).
Here we measure an even lower correlation length for galaxies
with z=[0.2,0.5] in the VVDS, coherent with their
very low intrinsic luminosity as discussed in Section~\ref{r0}.

\begin{figure}
   \centering
      \includegraphics[width=9cm]{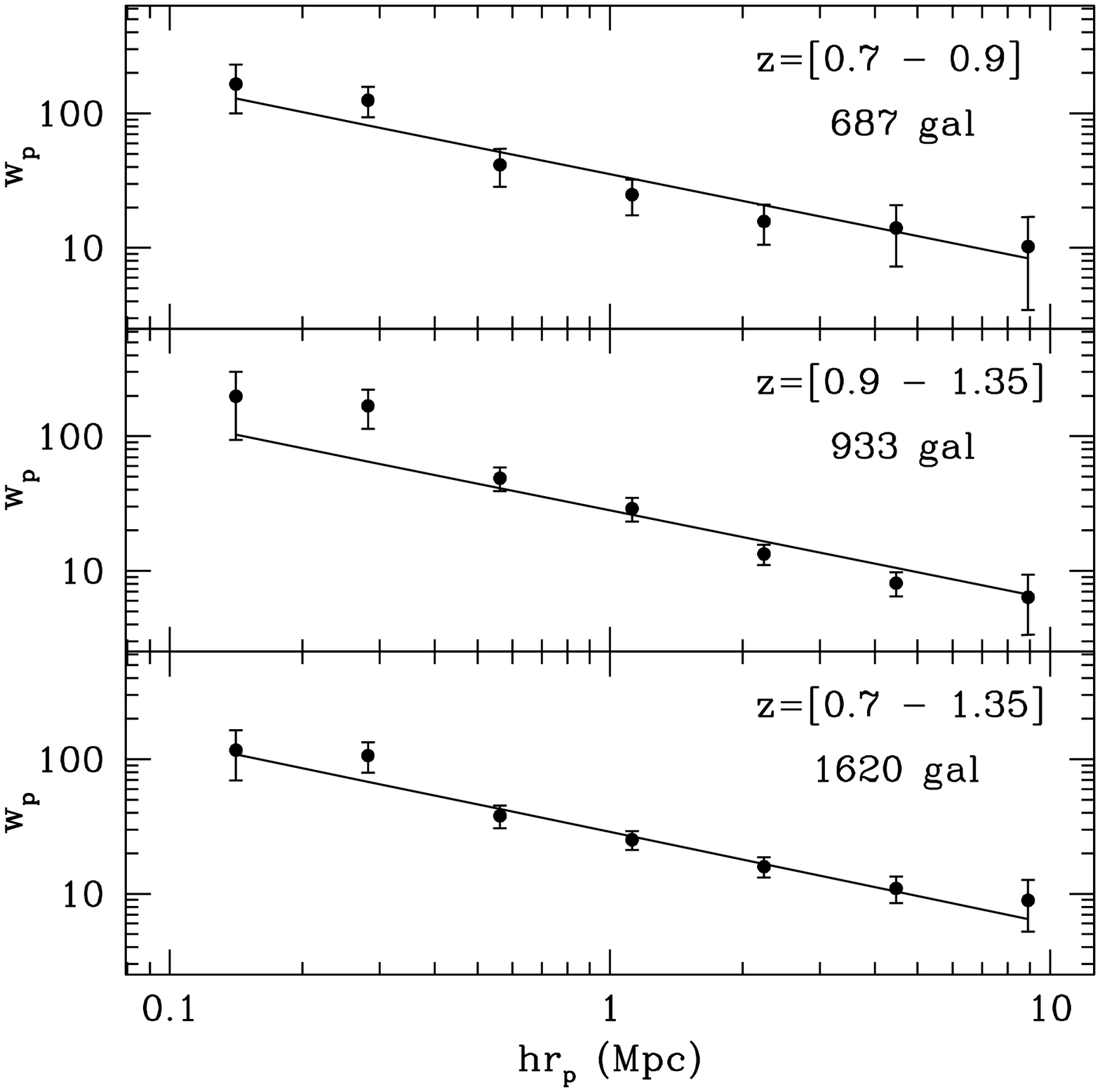}
      \caption{Correlation function computed from the VVDS
sample, applying the same color-magnitude selection
function as for the DEEP2 survey \citep{coil}: (bottom) the full
sample  in the redshift range [0.7,1.35], (top two panels) the
lower [0.7,0.9] and higher [0.9,1.35] redshift samples. The 
slope has been fixed to the same slope as measured in the
DEEP2 to ease $r_0$ comparison.
              }
         \label{deep2}
   \end{figure}

\begin{figure}
   \centering
      \includegraphics[width=9cm]{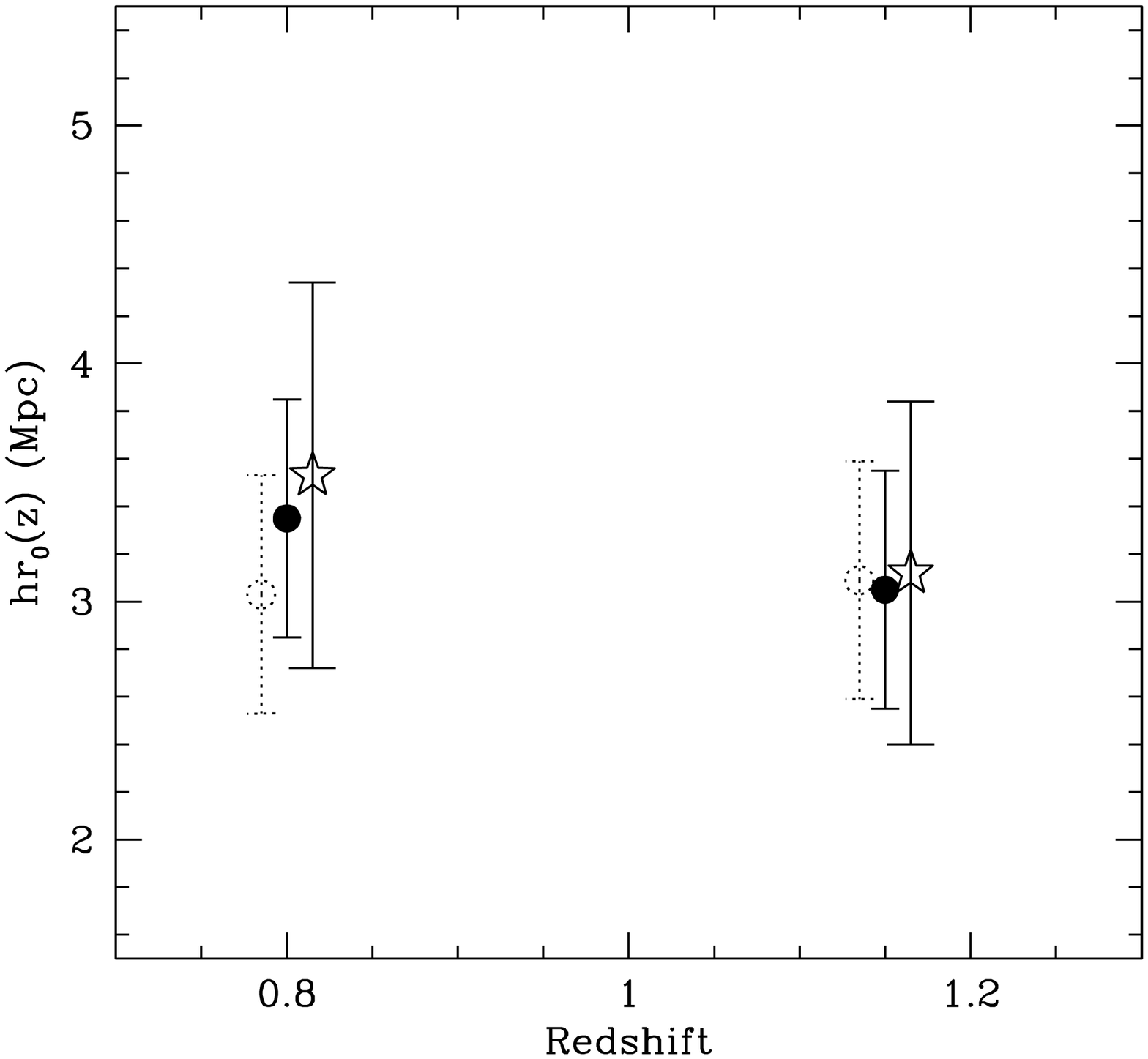}
      \caption{Comparison of the correlation length $r_0(z)$ (comoving)
measured in the VVDS-Deep survey 
applying a color-color and magnitude
selection as in  the DEEP2 survey, and the DEEP2 survey
measurements \citep{coil}: VVDS-02h, with  $\gamma=1.66$ fixed to the
DEEP2 value: filled circles; $r_0$ and $\gamma$
fitted simultaneously: open-dot circles, DEEP2 measurements: open stars.
Data points have been shifted slightly along the redshift axis to avoid
overlap.
              }
         \label{r0zdeep2}
   \end{figure}

In order to compare our measurements at higher redshifts
to the measurements
of the DEEP2 survey \citep{coil}, we have restricted our data
by applying {\it a-posteriori}
the same color and magnitude selection function as the DEEP2 survey
applied a-priori to pre-select its spectroscopic targets
in the redshift range $0.7-1.35$. With this ``DEEP2'' selection function
($R_{AB} \leq 24.1$, $B-R \leq 2.35(R-I)-0.45$, $R-I\geq1.15$ or
$B-R\leq 0.5$)
only 59\% of the VVDS I-band magnitude-limited sample galaxies are 
selected (in the desired redshift range).
The correlation function $w_p(r_p)$, and the
corresponding $r_0(z)$ and $\gamma(z)$ 
of this sample are shown in Figure \ref{deep2}. We find that
$r_0=3.51\pm0.63$ h$^{-1}$ Mpc for the full z=[0.7,1.35] sample,
compared to $3.19\pm0.51$ by \citet{coil}. Separating in
the same redshift bins, and setting the slope to
$\gamma=1.66$ as measured in DEEP2, we find that 
$r_0=3.37\pm0.67$ h$^{-1}$ Mpc
in z=[0.7,0.9], $r_0=3.05\pm0.47$ h$^{-1}$ Mpc in z=[0.9,1.35], 
while \citet{coil} find  $r_0=3.53\pm0.81$ h$^{-1}$ Mpc and 
$r_0=3.12\pm0.72$ h$^{-1}$ Mpc, respectively, as shown
in Figure \ref{r0zdeep2}. Given
the relative uncertainties of both surveys, these results 
are therefore in excellent agreement. 
The correlation function for the DEEP2 survey therefore 
differs from the correlation function of the complete VVDS sample
because of the different selection function applied that
selects galaxies brighter than in the 
VVDS and excludes galaxies relatively blue in R-I 
($R-I<0.7$) but still red
in B-R ($B-R>0.6$), resulting in a larger 
correlation length. 

\section{Discussion: evolution of the clustering length $r_0(z)$ from $z\simeq2$}
\label{r0}

\begin{figure*}
   \centering
      \includegraphics[width=12cm]{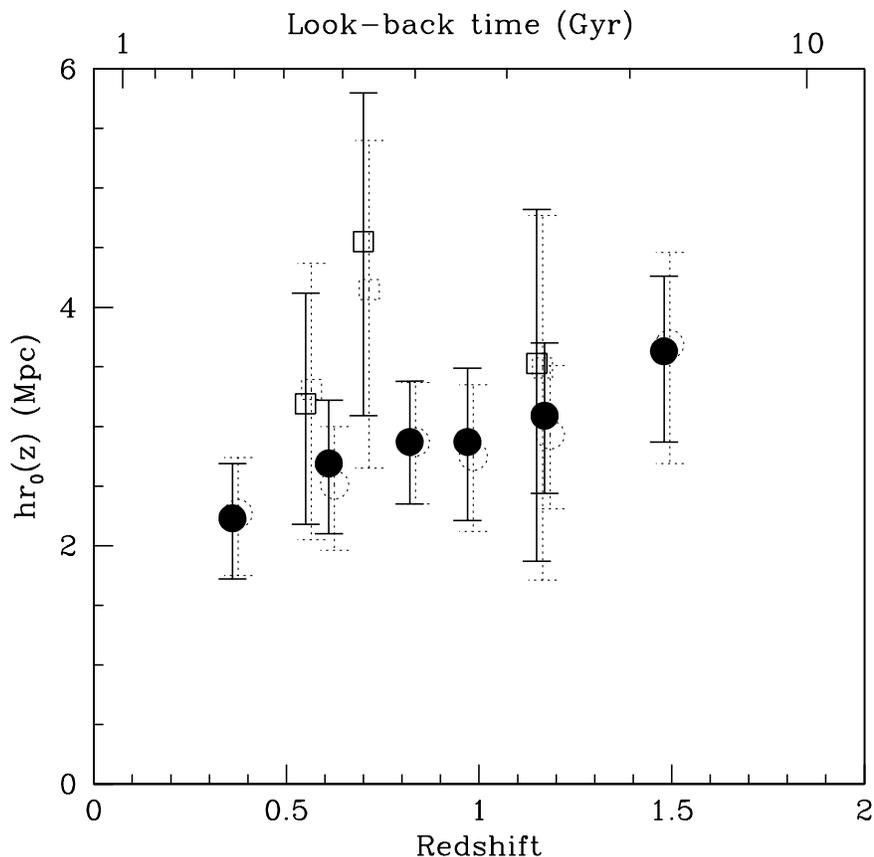}
      \caption{
Evolution of the correlation length $r_0(z)$ (comoving) from the VVDS.
Black circles are for the VVDS-02h data, open squares are for the VVDS-CDFS
data. The doted points indicate the measurements
when setting the slope to the mean $\gamma=1.76$ measured
in the range z=[0.2,1.3].
Associated errors have been computed from the fitting of $w_p(r_p)$
and associated errors drived from 50 mock galaxy samples from
Galics (see text). The
VVDS-CDFS measurements are slightly higher than in the VVDS-02h,
but remain compatible within the measurement errors. The VVDS-CDFS
measurement at z=0.7 include a strong over-density of more than
130 galaxies \citep[see][]{lefevre04b}, making the correlation
function higher in this bin.
              }
         \label{r0z}
   \end{figure*}

The evolution of the clustering length $r_0(z)$ from the VVDS first 
epoch ``Deep'' sample of galaxies with $17.5 \leq I_{AB} \leq 24$ 
is presented in Figure \ref{r0z}. The redshift of each bin is computed
as the mean of the redshifts of galaxies in each bin.
We find that $r_0$ is roughly constant or possibly
slightly increasing, within our measurement
errors, as  redshift increases,
with a low value of $r_0=2.35_{-0.37}^{+0.36}$ \hmpc\ for $z=[0.2,0.5]$, 
to $r_0=3.03_{-0.56}^{+0.51}$ \hmpc\ for $z=[1.3,2.1]$. 

Ideally, one would like to follow the evolution of the
clustering of the mass in the universe, translated into
the gravitational growth of structures. To access to this
measurement from the correlation
length of galaxies requires understanding of the evolution
of galaxies as complex physical systems, 
and to be able to relate the populations of galaxies observed at different
redshifts as descendents of a well-identified original population
of matter halos at early epochs. Although the VVDS
I-band magnitude selection is the minimum selection bias one can 
impose selecting a distant sample, this faint magnitude limit
leads to a broad redshift coverage, and to a population mix in the
VVDS sample that changes with redshift as described in Section 2.2.
Over the redshift range considered here, the
luminosity function of galaxies in the VVDS is strongly
evolving up to $\sim1.5-2$ magnitudes \citep{ilbert}. 
The interpretation of the evolution of $r_0(z)$
is therefore not direct. 

In the lower redshift bin $z\leq0.5$, we find a correlation
length smaller than any other population of galaxies observed today. 
The mean absolute luminosity of the
low redshift sample is $M_{B_{AB}}=-17.5$ with a significant
number of objects fainter than $M_{B_{AB}}=-16$ (Figure~\ref{MBz}),
while bright galaxies are under-represented 
due to the small volume available at these low redshifts.
Not surprisingly given the faint $I_{AB}=24$ cutoff in the VVDS, 
this makes it the faintest galaxy population
for which 3D clustering has ever been probed at low
redshifts, about 1.5 magnitude fainter than the 2dFGRS sample
with a mean $M_{b_{J}}=-17.98$ equivalent to 
$M_{B_{AB}}=-18.9$ \citep{norberg}. 
This low clustering of the low luminosity population of galaxies 
measured in the VVDS is roughly consistent with 
an extrapolation to fainter luminosities of the trend 
for a lower correlation length as galaxies become fainter,
as observed locally in the 2dFGRS and SDSS surveys 
\citep{norberg,zehavi02}, and at intermediate
redshifts in the CNOC2 survey \citep{sheperd}.

As redshift increases, the VVDS probes  
more of the bulk of the general population of galaxies.
Galaxies sampled at redshifts
$0.5-1.1$ become increasingly similar in luminosity and
color to the population sampled by the low
redshift 2dFGRS and SDSS surveys. As
can be seen from Figure~\ref{MBz}, at $z\simeq 1$ we
measure the clustering of galaxies with a mean absolute magnitude
$M_{B_{AB}}=-18.5$, after taking into
account $\sim1$ magnitude of luminosity brightening
at $z\simeq1$ \citep{ilbert}, thus comparable
to the bulk of the galaxies probed by 2dF and SDSS at $z\simeq 0.1$.  

Over the redshift range z=[0.5,1.1],
we find a slightly increasing correlation length
$r_0\simeq2.2-2.9$ h$^{-1}$ Mpc. For a similar, blue-selected, population of 
galaxies at
$z\simeq 0.15$, the 2dFGRS finds $r_0 \simeq 4.3-4.6$ h$^{-1}$ Mpc
\citep{norberg}, using the same technique we use here.  Note that at a
redshift similar to the 2dFGRS, the SDSS tends to sample a different mix of
morphological types, due to its red-based selection, and for this
reason measures a larger correlation length, $r_0 = 6.14 \pm
0.18$ h$^{-1}$ Mpc (\citet{zehavi02}; see \citet{hawkins} for discussion).
Our results
therefore seem to indicate that the amplitude of the correlation
function of galaxies which would have luminosities $M_{B_{AB}}=-19.5$ 
after correction for luminosity evolution is
about 2.5 times lower at $z=1$ than observed locally by the 
2dFGRS. 

While we know that galaxies with similar luminosities at different
redshifts may not be tracing the underlying dark matter
in a similar way (see \cite{marinoni}), 
this result is in qualitative agreement with the expectations of
the simple gravitational growth of primordial fluctuations
(see e.g. Figure 5 of \cite{weinberg}).  

In the highest redshift bins z=[1.1,1.3] and z=[1.3,2.1], 
we are measuring the correlation function
of the brightest $M_{B_{AB}} \leq -19.5$ galaxies,
with a mean of $M_{B_{AB}}\sim-20.5$ and
$\sim-21.0$ respectively. We observe that
the correlation length increases compared 
to the measurements in the range z=[0.5,1.1], up to $r_0\simeq3$ h$^{-1}$ Mpc
at z=1.2 and $r_0\simeq3.6$ h$^{-1}$ Mpc in the highest redshift bin
to $z\simeq2$.
The luminosity function of galaxies at these redshifts
shows a marked evolution, equivalent to an increase in
luminosity of $\sim1.5$ and $\sim2$ magnitudes
in z=[1.1,1.3] and z=[1.3,2.1], respectively \citep{ilbert}. 
These galaxies will therefore be expected to have a mean
$M_{B_{AB}}\sim-19.5$ at low redshifts after evolution. Again, comparing
the clustering observed in the VVDS to the clustering length
of $r_0\simeq5$  h$^{-1}$ Mpc measured for $M_{B_{AB}}=-19.5$ in the 2dFGRS 
\citep{norberg}, we find that the clustering amplitude
has increased by a factor $\simeq2.4$ from $z\sim1.3-1.5$
to $z=0$.

The evolution of the correlation length observed in our data is
in broad agreement with the results of computer simulations
of galaxy formation and evolution. In their SPH simulation,
\citet{weinberg} find that the clustering length of galaxies decreases
from $r_0\sim4.2$ h$^{-1}$ Mpc at $z=3$ to a minimum
$r_0\sim3.0$ h$^{-1}$ Mpc at $z=1.5$, then increases again to
$r_0\sim4.0$ h$^{-1}$ Mpc at $z=1$.  Note, however, that these
predictions refer to the correlation length of the same class of
galaxies, corresponding to $M>5 \times 10^{10} M_{\odot}$, ideally
followed at different redshifts.  In a real, magnitude-limited
observation, as we have discussed, this effect is related to the
different range of luminosities sampled at each redshift, and the
changing clustering strength at different luminosities. 
Besides the simulated clustering of galaxies,
the dark matter correlation
length is expected from theory and N-body simulations
to drop steeply with increasing redshift, from
$r_0\sim5$ h$^{-1}$ Mpc at z=0 to $r_0\sim1.8$ h$^{-1}$ Mpc at $z=1.5$
(see e.g. \citet{weinberg}). In comparison to these predictions, our
data clearly show that the clustering evolution of galaxies
does not follow the predicted trend for dark matter.

\section{Conclusions}

We have computed the
evolution of the correlation function $\xi(r_p,\pi)$ and its
integral along the line of sight $w_p(r_p)$, from the VVDS 
first epoch ``deep'' survey. The VVDS contains a large
spectroscopically-selected sample of 7155 galaxies representative
of the global galaxy population in the redshift range z=[0,2.1], in
a large $0.61 $deg$^2$ total area. 
The correlation length $r_0$ is observed to be low, 
$r_0\simeq2.2$ h$^{-1}$ Mpc, for the low
redshift $z \leq 0.5$ population, indicating the low
clustering of the very low
luminosity population sampled in this redshift range. Over the
redshift range z=[0.5,1.1], the correlation length of the population
of galaxies, with a luminosity range comparable to the lower redshift
2dFGRS and SDSS, stays roughly constant with $r_0\simeq2.8$ h$^{-1}$ Mpc.
At the highest redshifts probed in this paper, z=[1.1,2.1], we find
that the correlation length increases slightly to
$r_0\simeq3.6$ h$^{-1}$ Mpc. 

After applying the same selection function as
in the DEEP2 survey, our results are found to be in excellent 
agreement with the results of \citet{coil}.
However, the significantly different DEEP2 selection function 
excludes up to 41\% of the galaxy population observed by the VVDS,
making the correlation length of the DEEP2 sample larger than
in the purely magnitude-selected VVDS. The VVDS results demonstrate 
that the clustering of the whole
galaxy population at the same redshift is indeed significantly lower,
with $r_0\simeq2.9$ in the VVDS vs.  $r_0\simeq3.5$  h$^{-1}$ Mpc
in the DEEP2,
further emphasizing the importance of our simple, purely
magnitude-limited selection function.

Our measurements clearly show that the correlation length 
evolves only slowly with redshift
in the range $0.5 \leq z \leq 2$, in 
a magnitude limited sample with $17.5 \leq I_{AB} \leq 24$. 
Taking into account the different
VVDS galaxy populations probed as a function of redshift, with
intrinsically brighter galaxies probed
as redshift increases, we find that
the clustering of galaxies at z$\sim1-2$ in the VVDS
is about 2.5 times lower in amplitude than for the 
galaxies probed by the 2dFGRS at z$\sim0.15$, 
for populations with similar absolute $M_B$ magnitudes.
This result provides unambiguous evidence for clustering
evolution. 

Our results are in broad agreement with
simulations accounting for both gravitational growth and baryonic
physics \citep{weinberg,benson}.  These  simulations 
show that the underlying dark matter correlation 
evolves strongly with redshift, as expected in a hierarchical growth
of structures.   Our observation that the clustering
of galaxies does not follow such a strong evolution  therefore fully supports
the model in which luminous galaxies at $z=1-2$ (or 
$9-10$ billion years ago) trace the emerging peaks of the 
large-scale dark-matter distribution and implies a strongly 
evolving galaxy bias. 

We will investigate 
the evolution of the galaxy - dark matter bias elsewhere \citep{marinoni}.
A more detailed study of the dependance of clustering on
luminosity and galaxy type will be presented in forthcoming
papers.

\begin{acknowledgements}
This research has been developed within the framework of the VVDS consortium
(formerlly VIRMOS consortium).\\
This work has been partially supported by the 
CNRS-INSU and its Programme National de Cosmologie (France),
and by Italian Research Ministry (MIUR) grants
COFIN2000 (MM02037133) and COFIN2003 (num.2003020150).\\
The VLT-VIMOS observations have been carried out on guaranteed 
time (GTO) allocated by the European Southern Observatory (ESO)
to the VIRMOS consortium, under a contractual agreement between the 
Centre National de la Recherche Scientifique of France, heading
a consortium of French and Italian institutes, and ESO,
to design, manufacture and test the VIMOS instrument.
\end{acknowledgements}

\end{document}